%% file: main.tex
\documentclass[sigconf,natbib=false]{acmart}
\usepackage{subcaption}
\usepackage{hyperref}
\usepackage{multirow}
\usepackage{amsmath,amsfonts}
\usepackage{algorithm} 
\usepackage{algcompatible} 
\usepackage{algpseudocode} 
\usepackage{mathtools} 
\usepackage[style=ACM-Reference-Format,backend=bibtex,sorting=none]{biblatex}
\setcopyright{acmcopyright}
\copyrightyear{2018}
\acmYear{2018}
\acmDOI{10.1145/1122445.1122456}

\acmConference[XXXX]{XXXX}{XXXX}{XXXX}
\acmBooktitle{XXXX}
\acmPrice{XXXX}
\acmISBN{XXXX}

\begin{document}
\let\oldReturn\Return
\renewcommand{\Return}{\State\oldReturn}

\DeclarePairedDelimiter{\ceil}{\lceil}{\rceil}
\DeclarePairedDelimiter{\floor}{\lfloor}{\rfloor}

\title{Scalable privacy-preserving cancer type prediction with homomorphic encryption}

\author{Esha Sarkar}
\affiliation{%
  \institution{Tandon School of Engineering\\New York University}
  \city{Brooklyn}
  \country{USA}
}
\author{Eduardo Chielle}
\affiliation{%
  \institution{Department of Electrical and Computer Engineering\\New York University Abu Dhabi}
  \city{Abu Dhabi}
  \country{UAE}
}

\author{Gamze Gursoy}
\affiliation{%
  \institution{Program in Computational Biology and Bioinformatics\\Yale University}
  \city{New Haven}
  \country{USA}
}

\author{Leo Chen}
\affiliation{%
  \institution{Department of Computer Science\\Yale University}
  \city{New Haven}
  \country{USA}
}

\author{Mark Gerstein}
\affiliation{%
  \institution{Program in Computational Biology and Bioinformatics\\Yale University}
  \city{New Haven}
  \country{USA}
}

\author{Michail Maniatakos}
\affiliation{%
  \institution{Department of Electrical and Computer Engineering\\New York University Abu Dhabi}
  \city{Abu Dhabi}
  \country{UAE}
}

\begin{abstract}
Machine Learning (ML) alleviates the challenges of high-dimensional data analysis and improves decision making in critical applications like healthcare. Effective cancer/tumor classification from high-dimensional genetic mutation data can be useful for cancer diagnosis and treatment, if the distinguishable patterns between cancer/tumor types are identified. At the same time, analysis of high-dimensional data is computationally expensive and is often outsourced to cloud services. Privacy concerns in outsourced ML, especially in the field of genetics, motivate the use of encrypted computation, like Homomorphic Encryption (HE). But restrictive overheads of encrypted computation deter its usage. In this work, we explore the challenges of privacy preserving cancer detection using a real-world dataset consisting of more than 2 million genetic information for several cancer types.
Since the data is inherently high-dimensional, we explore smaller ML models for cancer prediction to enable fast inference in the privacy preserving domain. 
We develop a solution for privacy preserving cancer inference which first leverages the domain knowledge on somatic mutations to efficiently encode genetic mutations and then uses statistical tests for feature selection. Our logistic regression model, built using our novel encoding scheme, achieves 0.98 micro-average area under curve with 13\% higher test accuracy than similar studies. We exhaustively test our model's predictive capabilities by analyzing the genes used by the model. Furthermore, we propose a fast matrix multiplication algorithm that can efficiently handle high-dimensional data. 
Experimental results show that, even with 40,000 features, our proposed matrix multiplication algorithm can speed up concurrent inference of multiple individuals by approximately 10x and inference of a single individual by approximately 550x, in comparison to standard matrix multiplication.
\end{abstract}





\maketitle
\input{intro2}

In a nutshell, for a private cancer detection solution to be practicable, it must have three properties: 1) comprehensive high-dimensional genomic analysis for high detection accuracy with a focus on explainability, 2) cryptographically secure privacy guarantees, 3) practical inference time and high throughput. In developing the privacy-preserving tumor prediction model, we list our contributions as follows:
\begin{enumerate}
    \item We develop a three-step privacy-friendly feature engineering methodology and demonstrate the importance of feature selection and encoding based on biological significance of genetic alterations and statistical tests. We discuss the \textit{predictive genes} from our findings and compare it to Gene Ontology enrichment analysis to understand the significance of certain genes in predicting a cancer type.
    \item We build 1) a machine learning model to counter the imbalance in the dataset, and 2) high performing binary models using the same encoding for each cancer type. We compare our methodology with the reported baseline and we achieve a $13\%$ increase in accuracy \cite{deep_gene,gdl}.
    \item We propose a fast matrix multiplication algorithm for high-dimensional matrices, specifically designed to implement HE-based
    privacy-preserving logistic regression model to ensure high performance. 
    \item We optimize our private cancer prediction methodology for low latency as well as high throughput.
    \item We implement our privacy-preserving cancer-prediction model using BFV, an FHE scheme, and demonstrate that we can perform privacy-preserving cancer prediction for $\approx$500 individuals under 1 minute and a single prediction in approximately 1 second. We compare the performance of our privacy-preserving model with an ML model implemented using standard matrix multiplication.
    \item We open-source our methodology and implementation to be used as-a-service. 
\end{enumerate}

We first discuss the related work on genomics-based disease detection and privacy preserving ML inference in section \ref{s:related} and then the background information about the dataset and HE in section \ref{s:prelim}. We describe our complete methodology including plaintext model development using encoding scheme and fast matrix multiplication in section \ref{s:methods} and the experiments on performance and comparison in section \ref{s:results}. We finally conclude in section \ref{s:conclusion}.

\section{Related work}\label{s:related}
The study of genomic data to understand the changes that led to cancer or the genomic alterations that happened as a result of cancer is of utmost importance for initial diagnosis, predicting stage, cancer growth, metastasis, treatment, drug response, and planning a path to recovery. Therefore, there have been several studies focused genomic dynamics on individual cancer types. Researchers from TCGA have delved into genomic and molecular characterization studies for several individual cancer types: Glioblastoma \cite{glioblastoma}, ovarian carcinoma \cite{ovarian_carcinoma}, lung \cite{lung}, endometrial carcinoma \cite{endometrial}, renal cell carcinoma \cite{renal}, urothelial carcinoma \cite{urothelial}. In 2012, TCGA launched a pan-cancer dataset collection i.e. coherent dataset collection over 12 tumor types each profiled using 6 different platforms: Reverse-phase protein arrays measuring protein and phosphoprotein abundance (RPPA), DNA methylation, microarray-based measurement of copy number,  single-nucleotide and structural variants mutation using whole exome sequencing, microRNA sequencing, and RNA sequencing and microarray gene expression analysis~\cite{pan_cancer}. This opened up several new avenues of cancer analysis using genomic data like identification of genes that drive carcinogenesis~\cite{driver_genes}, studies on the metastatic nature of cancer types~\cite{metastatis}, and using genomics for precision medicine \cite{precision}. All of these studies that focus on detecting the cancer type analyze the genomic data to help develop biological intuition towards understanding a tumor type. Therefore, we also perform the encoding of genomic mutation data based on biological intuition to retain this wealth of semantic information. 

Prediction of cancer type using machine learning on genomic data has been of interest in the last few years because of the possibility of computation on the huge volume of genomic data. Researchers have aimed at finding information about cell of origin for all the 33 different tumor types \cite{similar_origin}. Jiao et. al. propose a deep learning-based framework to predict 24 tumors of unknown primary site by analyzing somatic passenger mutations. Machine learning has also been used to analyze genomic data for immunotherapy for pan-cancer datasets \cite{inhibition}. Deep learning has also been used for the prediction of tumor type using gene filtering \cite{gdl} and mutation frequency, sparsity reduction, and cluster gene filtering \cite{deep_gene} as pre-processing steps. Another approach to finding/understanding useful features is through auto-encoders, as proposed for individual cancer detection \cite{encoder_liver, encoder_breast} or for classification of 40 different cancer types achieving an area under curve between 0.54-0.97 for individual types~\cite{auto}. Our work also revolves around efficient tumor detection but we observe that a combination of biological intuition and statistical tests is required for a higher performance in tumor type detection. 

Amidst the growing privacy concerns for sensitive data, private inference has been a topic of interest in recent years. Cryptonets \cite{cryptonets}, one of the earlier research articles on generic HE-based private inference, implemented DNN by approximating non-linear layers with some lower order polynomial terms. Cryptonets, like many HE algorithms, are also optimized for throughput, predicting class labels for 8192 images together. The performance was improved by several other solutions with Gazelle~\cite{gazelle}, a low latency framework, being able to achieve state-of-the-art performance of 800 ms for one inference. HE-based ML/DNN implementations focus on lowering the cost of non-linear functions like ReLUs as the feature space (and thus, the cost of matrix multiplication) is smaller. For example, the largest inputs considered by the solutions is from the CIFAR dataset, with 3K features, whereas our model needs a minimum of 30K features. In this work, we focus on a different problem, where we require high number of features, high throughput, as well as low latency.

Private inference on genomic data is a challenging problem because of the nature of data. Several challenges in HE-based genome privacy were developed in iDash (integrating Data for Analysis, 'anonymization,' and SHaring) like private statistic calculation~\cite{idash15}, privacy-preserving querying~\cite{idash16}, and logistic regression training on a dataset of 18 features and 2 labels~\cite{idash17}, each focusing on a different bottleneck of genome privacy road-map. iDash19 and iDash20 focused on private inference challenges. We observe that rigorous feature engineering is often performed on the original data to reduce the number of features to a bare minimum, from 16K features to 10-40 features \cite{access_imputation,ultrafast}. Authors have performed genomic analysis using several different libraries and types of FHE like CKKS, BFV, and TFHE using different models like SVM, LR, shallow DNNs~\cite{ultrafast}, and also using partial homomorphic encryption \cite{access_imputation}. Naturally, as the number of features are reduced by three orders, there is a reduction in test accuracy, which is justified with performance vs accuracy trade-off. On a different application (genome wide association studies involving a large number of individuals), researchers have also explored approximation techniques in logistic regression algorithms using semi-parallel implementation \cite{idash18} improving evaluation time by 30\% ($\approx$ 6 hours on a single node). 
Our privacy preserving inference methodology is specifically directed to problems where a large number of features is indeed required.

\section{Preliminaries}\label{s:prelim}
\subsection{Dataset}\label{ss:dataset}
We use the cancer classification dataset from iDASH 2020 competition Task I \cite{idash20} that was collected for private tumor classification. This data is curated from a centralized database The Cancer Genome Atlas (TCGA) ~\cite{TCGA} using patients from 11 different cancer types. 
As noted in section \ref{s:related}, several subsets of TCGA resulted in different types of studies. In our work, we study the impact of somatic mutations on prediction of cancer. Our dataset consists of two types of somatic alteration information (maybe considered as two subsets of features): Single-Nucleotide Variations (SNVs) and Copy Number Variations (CNVs) on protein-coding genes.  In the SNV subset, four different characteristics are given for each somatic SNV of a gene. These characteristics represent the chromosome location, denotes whether the mutation is a single-nucleotide polymorphism, and the effect of the mutation (using two different measures). The effect of the mutation is calculated using Ensembl Variant Effect Predictor (VEP)~\cite{vep} and is reported in two ways: 1) A mutation can be considered as one of the following categorical values; high, moderate, modifier, and low, followed by a real number denoting the impact of the mutation. 2) A mutation can qualitatively be denoted as tolerated or deleterious, based on Sorting Intolerant from Tolerant (SIFT) pathogenicity prediction. All of this information reflects the importance of a mutation, i.e. VEP scores help transform an observation of a mutation to its possible impact in development of the tumor. VEP scores help in developing the biological intuition for our feature engineering methodology, which is required as this subset of SNV features contains 2,044,328 somatic mutation rows.
In the copy number subset, each gene for each sample (patient) is given a copy number value depending on whether there has been a change from their parents' genes: 0 for no alteration, 1 or 2 for duplication, and -1 or -2 for deletion from one or both the parents, respectively. For each sample, there are 25,128 genes, and thus, there are 25,128 features for each sample.
The dataset comes from 2713 patients belonging to 11 different cancer types. The dataset is unbalanced with the maximum number of patients belonging to Bronchus and Lung ($23.51\%$). We randomly select $80\%$ of the dataset for training and the rest is reserved for testing maintaining the distribution of data, i.e. the distribution of training and test data for each label remains the same.

\subsection{Homomorphic encryption}
Homomorphic Encryption (HE) is a type of encryption that allows for computation on encrypted data without decryption. Let us consider a function, $f(.)$ operating on plaintext operands $p_1,p_2$, and the equivalent function $f_{enc}(.)$ operating on the corresponding ciphertexts $c_1,c_2$, such that $c1 = Enc(p_1)$, and $c2 = Enc(p_2)$, where $Enc(.)$ is the encryption function. Then, the computation of the function $f(.)$ on plaintext operands $p_1,p_2$ is the decryption of computation of the function $f_{enc}$ on ciphertexts, i.e. using HE, we can say that $f(p1,p2)=Dec(f_{enc}(c_1,c_2))$, where $Dec(.)$ is the decryption function. Depending on the type of computation possible in an encryption scheme, there are several types of HE schemes. 

For linear models with unencrypted weights, Partial Homomorphic Encryption (PHE) schemes like Paillier \cite{Paillier} can be used. Nevertheless, encryption and decryption operations, which consist of modular exponentiations, hinders the performance of ML models with larger inputs or outputs. In addition, although it is possible to encode several plaintext into a ciphertext in Paillier for certain applications, the density of plaintexts per ciphertext is much lower than in Somewhat Homomorphic Encryption (SHE) or Fully Homomorphic Encryption (FHE).

A better approach comes from using SHE/FHE schemes like BFV (Brakerski/Fan-Vercauteren) \cite{bfv} or CKKS (Cheon, Kim, Kim, Song) \cite{ckks}. CKKS enables fixed-point arithmetic and it is the standard choice for ML applications. During computation, CKKS drops the lower bits of the plaintext after each operation, reducing the precision of the result.
Conversely, BFV works on integers (modular arithmetic), where we can emulate a fixed-point arithmetic by scaling up the double-precision floating-point number into integers. Similarly to CKKS, there is a limitation for how much precision a BFV ciphertext can provide. However, since it computes on modular arithmetic, we can use the Chinese Remainder Theorem (CRT) to break our values into several smaller values, each one under unique modulus coprime to all other moduli. Each smaller value is then encrypted under a different key.
\section{Threat model} 
\begin{figure}
    \centering
    \includegraphics[scale=0.5]{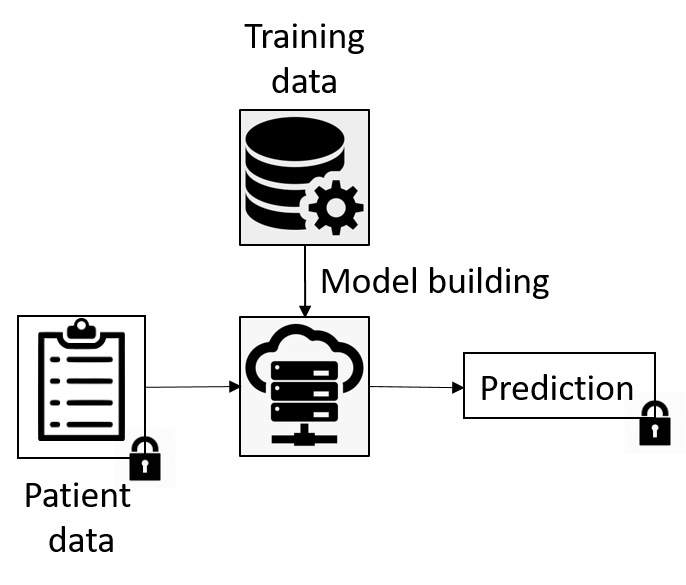}
    \caption{Our threat model for private inference.} 
    \label{fig:threat_model}
\end{figure}

In this work we focus on privacy-preserving inference. In our threat mode, the training dataset is public and is comprised of individuals who have agreed to share their data. The Cancer Genome Atlas database is one such example \cite{TCGA}. This database catalogues genomic information, prognosis, diagnosis, on top of personal information like age, gender, race, and ethnicity of the individuals who are identified as case numbers. We do not aim to protect the training dataset, rather use it to build models for cancer prediction. However, when a new patient wants to test for cancer using their genome, their data must be protected. In our threat model, thus, the training is not privacy-preserving, but the inference is private. Fig. \ref{fig:threat_model} summarizes our threat model.

\section{Methodology}\label{s:methods}
Since computations in the encrypted domain are expensive, private inference on any type of should prioritize towards less number of computations. This translates to a low number of features and smaller ML models (for example, SVM or logistic regression instead of deep networks). Our methodology can be divided into two parts. In the first part of our methodology (subsection \ref{ss:feature_encoding}), we focus on making our dataset compact encoding mutations and reducing the number of features with biological intuitions and statistical tests. This reduces the number of features from over 2 million to 43K. In the second part of our methodology (subsection \ref{ss:mat_mul}), we propose a matrix multiplication algorithm , particularly catered towards implementing a faster version of privacy-preserving logistic regression-based cancer inference.

\subsection{Somatic mutation encoding}\label{ss:feature_encoding}
For the cancer prediction to be correlated to both CNV and SNV information, the CNV and SNV features can be concatenated together, which can be used to train an ML model. CNV subset has 25K features. But the SNV subset corresponds to over 2 million mutation rows, which may equate to over 2 million features if each of these mutations is analyzed separately. The concatenated dataset, thus, consists of more than 2 million features. Therefore, instead of representing a mutation as a feature, we represent a gene as a feature with encoded mutation as the value of that feature. 
However, this approach faces the challenge of compacting mutation information of over 2 million data-points to 25K data-points (corresponding to 25K genes). 
\subsubsection{Step 1: Gene/feature selection using SNV frequency}
Previous studies~\cite{deep_gene, gdl} on cancer detection using somatic mutations observed that the frequency of mutation of a gene correlates to higher prediction accuracy. The genes with higher number of mutations are more likely to develop a cancerous tumor. This makes sense because if there are more mutations in a gene, then the expression of that gene is likely disrupted through the mutations. There is also a correlation between the frequency of mutations in a patient cohort and the likelihood of observing the cancer in that cohort. This also makes sense because if a mutation is observed in a large number of patients with same cancer type diagnosis, then that mutation is likely either correlated or causal to that cancer type. Therefore, as the first step of feature selection, we choose genes with higher number of mutations. But each cancer type corresponds to a higher mutation in a different gene.
We first rank the genes based on their SNV frequency in the patient cohort for each cancer type by also taking into consideration genes with more than one SNVs on them. We then combine the ranked genes from different cancer types and finally select the top 10,000 genes with highest SNV frequency for each cancer type as our features. This step also ensures removal of genes with low SNV frequency and reduce the dimensionality of our feature space. With all cancer types combined, we have a total of 18,606 genes. 
\subsubsection{Step 2: Encoding scheme}
Encoding techniques, both for the feature vectors and the target variables, ease in training towards better accuracy. Efficient encoding techniques have been proven to result in better performance in genomic classification tasks as well \cite{encoding}. 
As mentioned in the dataset section, each SNV on a gene is represented by multiple characteristics. This information need to be meaningfully merged with the CNV information of each gene. We explore the following encoding schemes based on a biological intuition as explained below:

\begin{enumerate}
    \item Using presence of an SNV on a gene: The genes selected using frequency are merged for all cancer types. In this encoding we aim to study if a particular gene (mutation) is highly correlated to a cancer type.
    We assign a binary value $[0,1]$ to each of genes of a patient to denote the presence or absence of one or more SNVs. This allows us to encode SNV information in a categorical manner. This binary value per gene is used as a feature.
    \item Using the type and impact of an SNV: As denoted in subsection \ref{ss:dataset}, the impact of mutation of a gene is calulated using VEP 
    For each SNV in the dataset, we have two types of measure: 1- a qualitative measure indicating whether an SNV has a tolerated or deleterious effect; 2- a quantitative real-value measure ($s_{i,j}$) representing the strength of the impact and the qualitative confidence with which the mutation can be attributed to the diagnosis. $s_{i,j}$ is the strength of the $i^{th}$ SNV in patient $j$. Each of the encoding scheme are given equidistant real values between 0 to 1. We also experimented with different ranges but did not find any improvement in test accuracy. We think that both of these measures are useful in classifying the tumor type. 
    We first encode the first measure by assigning values for  [deleterious, deleterious (low confidence), tolerated (low confidence), tolerated] as [1.0, 0.75, 0.5, 0.25] for each $m_{i,j}$. The reason for this encoding is that a detrimental effect is given the highest value in cancer prediction and similarly a tolerated effect is given the lowest feature value. Either of the effects, when estimated with lower confidence are given lower effect. $m$ is the effect of the $i^{th}$ SNV in patient $j$. We then combine this encoding with the second measure as $m_{i,j} \times s_{i,j}$. The final effect values of SNV impact of a gene is the summation of the impacts of all the SNVs on that gene, if there is more than one SNVs. The resulting value per gene is used as a feature.

    In addition to the strength ($s$) of an SNV, the qualitative confidence of the effect of an SNV on a gene is also as a categorical variable with values as [high, moderate, modifier, low]. We encode them as [1, 0.4, 0.7, 0.1] since intuitively we want to assign a higher importance to a high mutation effect. A gene with no mutation is assigned 0 effect. Similar to the previous feature, for a gene with multiple SNVs at different locations, the values are added to finally represent the effective value of all the SNVs in a gene. The resulting value per gene is used as a feature (Figure 1).
    \item {Using CNV of a gene}
CNV of a gene is represented as integers between -2 and 2, indicating if both, one, or no copy of the gene is deleted or duplicated. We scale these values to integers between 0 and 4 as statistical feature selection methods (for example $\chi^2$ test) often require positive values. Similarly, the resulting value per gene is used as a feature.

Overall, the top 10,000 genes per cancer type are merged into a total of 18,606 genes. For each patient, 18,606 genes with their associated SNV encoding are concatenated with 25,128 genes with their CNV information. In total, we have 43,734 \textit{features} which undergo the following statistical tests. 
\end{enumerate}




\subsubsection{Step 3: Feature selection using $\chi^2$ test}
The previous steps of feature selection incorporate biological intuition. In this step, we explore statistical tests for evaluating feature importance. 
Statistical methods like $\chi^2$ test do not only inform the feature importance but also help reduce dimensionality of the feature space enabling faster computation (both during training and inference). It has previously been used in genetic information based disease prediction studies \cite{chi_cancer}. We choose $\chi^2$ test as a feature selection metric since we achieve the best possible accuracy when compared to feature selection with mutual information or f-score statistics (as reported in Section \ref{s:results}).
To decide if a feature is independent of the target label (i.e type of cancer), we perform the $\chi^2$ test where we calculate $\chi^2$ value of each feature with respect to the target variable. The $\chi^2$ value of a feature is given as $\sum \frac{(O_i-E_i)^2}{E_i}$ where $E$ represents the expected value, $O$ represents the actual output and $i$ represents each instance of a $\chi^2$ test between a feature and a target. Note here that, the expected values of a variable is calculated using the distribution of feature values.

We run the $\chi^2$ test on all genes (CNV and SNV concatenated together) and sort the features decreasing order of $\chi^2$ values. 
The top $n$ features are selected and are used to train a classification model. We also use this step to analyze the selected genes and their relative \textit{importance} in cancer type prediction in Section \ref{ss:predictive}.

\subsection{Privacy-preserving cancer inference}\label{ss:mat_mul}
\subsubsection{Model selection for cancer prediction}For the selection of the ML model that can correlate encoded somatic mutations to cancer type, we train a variety of ML models, while considering the difficulty of their implementation in HE domain.
In plaintext, we experiment with Support Vector Machine (SVM) (with radial basis function, polynomial and linear kernels), logistic regression, and Deep Neural Networks (DNNs with two fully connected hidden layers with relu activation) possible classification models. 
We start with top 1,000 features selected by $\chi^2$ test and increase the number of features by 1,000 in each iteration. In our search for best-performing model, we train models using different number of features, different statistical tests for feature selection, different kernels (if applicable), with several regularization techniques, and with different optimization techniques cross-validated over 5 folds. We report the best performing models in Section \ref{s:results}.

\textbf{Measures to reduce overfitting:} We find that the logistic regression model performs best for several encoding schemes with the best test accuracy of $83.61\%$. In logistic regression, the probability that a sample $x$ belongs to class $k$ is given by $P(y=k|x)=\frac{e^{z_k}}{\sum^K_{l=1}e^{z_l}}$ where $z$ is the linear combination of coefficients of the form $\beta_j$. However, the number of features is much larger than the number of samples since each sample has effectively $\approx$ 43,000 features from CNV and SNV, whereas we have a total of 2,173 samples. Therefore, to avoid over-fitting in this high-dimensional setting, we introduce a Lasso ($l1$) penalty \cite{lasso} to the logistic loss function during training such that the features that are unlikely to contribute to the prediction are penalized and weighted zero. Therefore, if the logistic loss function is given by $L(\beta_j)$ where $\beta_j$ represents the coefficients of the features, the loss function after Lasso penalty in the Lagrangian form becomes $L(\beta_j) - \lambda \sum_j^n|\beta_j|$ which is minimized during training. Further, we use k-fold cross-validation on the training set ($k=5$) as another way to prevent the model from over-fitting to the training data. We iterate for 10,000 epochs to converge to a prediction model. 

\textbf{Metrics to detect problems of unbalanced data:} High test accuracy on unbalanced datasets (with a higher percentage of samples from a particular label) can give a false sense of performance as a random guess (of the label with the highest number of samples) may also result in a high accuracy. For a holistic performance evaluation of our classifiers, we plot Receiver Operating Characteristics (ROC) Curve and report the individual area under curve for each class and the Micro-average Area Under Curve (MAUC) for the classifier. Although our dataset is unbalanced for 2 classes, still we report both test accuracy and ROCs for all the classes.

\textbf{Binary models:} Our cancer type prediction model uses the somatic genetic information to predict the cancer type from 11 classification labels. We also build models to predict each cancer type separately (like specific models in \cite{gdl}). These models are supplementary to our main prediction model and focus on one type of cancer. The features for this single cancer type model are chosen following the steps described above and the classifier is trained using a binary label: 0 for the all of the other cancer types and 1 for the cancer type of interest. We create separate models for each of the 11 cancer types and call these models as binary models since the prediction is converted into a binary classification task. 

It should be noted here that the \textit{binary} classification ability represented by the ROC curves of individual diseases (in our main prediction model) and the binary models for individual diseases are different because of the feature selection steps. In our binary models, the genes important to a specific disease are selected. However, for our main prediction model, the genes which are cumulatively important for all the 11 labels, are selected. Hence, we report both the analyses in section \ref{s:results}. 

From analyzing the nature of genomic mutation data and the trends in accuracy (details in section \ref{s:results}), we observe that regardless of the ML model selected, the matrix multiplication would involve high-dimensional matrices (Dot product between weights and several thousands of features is common for all the ML models explored in our work). Therefore, following standard matrix multiplications would require a large number of multiplications corresponding to this high-dimensional dataset. The private cancer prediction methodology is characterized by a private inference protocol proposed in subsection \ref{sss:protocol} and then the fast matrix multiplication methodology, crux of the private ML algorithm, is proposed in subsection \ref{sss:mat_mul_algo}.

\subsubsection{Private inference protocol}\label{sss:protocol}
\begin{figure}
    \centering
    \includegraphics[scale=0.5]{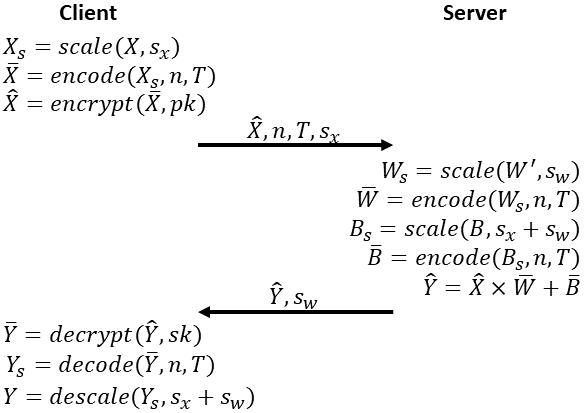}
    \caption{Overview of proposed inference protocol. The accent $\bar{\cdot}$ represents encoded data, while $\widehat{\cdot}$ denotes encrypted data.}
    \label{fig:protocol}
\end{figure}

Fig. \ref{fig:protocol} shows the overview of our inference protocol. It consists of input encoding and encryption, weight and bias encoding, computation, decryption, and decoding.
The client starts with a $|X| \times f$ matrix $X$ containing the input values represented with double-precision floating-point numbers, where $|X|$ is the number of inputs and $f$ is the number of features. The values of matrix $X$ are multiplied by a scaling factor $2^{s_x}$ in order to be converted into integers, a requirement of the BFV encryption scheme. This effectively converts our HE operations into fixed-point arithmetic.
The scaled matrix of inputs $X_s$ is then encoded into a matrix of polynomial plaintexts $\bar{X}$, where each polynomial contains $n$ coefficients. We pack $n$ features of each row into a polynomial. This leads to an encoded matrix of dimensions $|X| \times \ceil{f/n}$.

It is worth noting that our model requires more precision than what can be represented in the plaintext modulus $t$. From experimental results, we determined that our inputs and weights require 14 bits of precision. Since our inputs are in the interval $0 \leq x < 2^8$, we set $s_x = 6$ to represent the inputs in 14 bits. Meanwhile, weights are in the interval $0 \leq w < 1$, which leads to $s_w = 14$.
Due to the fixed-point arithmetic, the biases must be scaled by $2^{s_x+s_w}$.
After the computation, the client will receive outputs that are scaled by a factor of $2^{s_x+s_w}$, like the biases, but that require $8 + s_x + s_w + \ceil{\log_{2}(f)}$ bits of precision for representation. For $f = 40960$, that translates to 44 bits.
This is above of what a secure BFV ciphertext with enough noise budget for our computation can support. To cope with that without hindering the accuracy of our model, we used the Chinese Remainder Theorem (CRT) to break our plaintext into a pair of smaller plaintexts, each one under its own modulus.
We define our plaintext moduli $T = \{t_0, t_1\}$ as $t_0 = 1073872897$, which provides 30 bits of precision, and $t_1 = 114689$, offering 16 bits.
This means that for every $n$ features encoded into a polynomial, we are actually encoding into a pair of polynomials, one with coefficient modulus $t_0$, and another with coefficient modulus $t_1$. For simplicity, we refer to this pair as plaintext polynomial.

The encoded matrix of inputs $\bar{X}$ is then encrypted with the client's public key $pk$. The encrypted matrix $\hat{X}$ is sent to the server together with public values $\{n, T, s_x\}$.
Afterwards, the server scales and encodes the transpose of the matrix of weights $W$, which has dimensions $f \times |Y|$, where $|Y|$ is the number of outputs. The transposition is a requirement of our computation. It packs several feature weights of an output in a plaintext polynomial, leading to an encoded matrix of weights $\bar{W}$ of dimensions $|Y| \times \ceil{f/n}$. Biases are encoded differently, each bias is encoded into a plaintext polynomial, filling all slots with its value. Finally, the server performs the matrix multiplication of encrypted inputs by encoded weights followed by addition of encoded biases $\hat{Y} = \hat{X} \times \bar{W} + \bar{B}$. The resulting matrix $\hat{Y}$ is returned to the client together with public value $s_w$.
The client simply decrypts, decodes, and descale $\hat{Y}$ with its secret key $sk$ and obtains the result of the inference in plaintext.

\subsubsection{Matrix multiplication algorithm}\label{sss:mat_mul_algo}

Our privacy-preserving matrix multiplication algorithm, optimized for implementation in HE, is displayed in Algorithm \ref{alg:mm}. It receives three arguments: The encrypted matrix of inputs $\hat{X}$, encoded matrix of weights $\bar{W}$, and polynomial degree $n$. Each row of $\hat{X}$ represents an input, while each row of $\bar{W}$ represents one output.
The computation of the dot product for each input is independent, making this algorithm highly parallelizable.
We start the dot product by performing the column-wise multiplication of each row of $\hat{X}$ with the each row of $\bar{W}$ and append the result for each row-row pair into a vector (lines 5-11).
Next, we add together all elements of the resulting vector (line 13) and execute $\log_2{n}$ ciphertext rotations and additions to finalize the dot product (lines 14-17).
This results in a ciphertext where all its slots contain the result of the dot product of the row-row pair.
In order to save memory and reduce communication time, we aim at packing several dot product results into a single ciphertext. For this, first we need to clear the ciphertext slots in all but one carefully chosen position. We do it by multiplying the resulting ciphertext $\hat{c}$ by a plaintext polynomial $\bar{p}$ with one at that specific position and zero in the remaining slots (lines 18-22).
Finally, we can compress the dot product results $\hat{R_0}$ by adding them together (lines 25-42). If there are more dot product results than slots in a ciphertext, i.e., $|\hat{X}| \cdot |\bar{W}| > n$, then ciphertexts are appended to the output vector $\hat{Y}$. Lastly, we return the result $\hat{Y}$ (line 43). We provide the mathematical representation of the algorithm in \nameref{s:appendix}
\input{algorithm}
\subsubsection{Approximation of non-linear function in private inference}\label{sss:method_approx}
Tumor prediction is a classification problem which we address using multinomial Logistic Regression (LR). An LR model is trained by reducing the logistic loss function. During inference, the probability that an input $(x \in \mathbb{R}^{1\times d})$, with $d$ features, belongs to a class $(k)$ is given by $P(y=k|x) = \frac{e^z_k}{\sum^K_{l=1} e^z_l}$ where $z=Wx+b$, $W \in \mathbb{R}^{K\times d}$ is the weight matrix, and $b \in \mathbb{R}^{d}$ is the bias. The predicted class $(k_p)$ is the class with the highest probability, i.e. $k_p=argmax({P(y=k_i|x)})$ where $ k_i \in K$. This non-linear logistic function is computationally expensive in HE; thus, we perform the following approximation for building an ML model that can be used for encrypted inference.
Since the logistic function is a monotonically increasing function, we can say that $P(y=k|x)$ for a class is higher if $z_k$ is higher, and since the predicted label depends on the relative probability values, the predicted label can also be calculated using $argmax({z_{k_i}})$. Therefore, during inference, a test input $(x_{test})$ needs to be multiplied with the weight matrix to get the final prediction, i.e. the predicted class $k_p = argmax(W \times x_{test}+b)$. Effectively, for efficient inference, the matrix multiplication between the test inputs and the weight matrix must be fast. Please note here that the size of $W$ is dependent on the number of features, i.e. the higher the dimension of an input, the larger is the $W$ matrix, and the more time-consuming is the matrix multiplication.

\begin{figure*}[]
    \centering
        \includegraphics[width=0.32\linewidth]{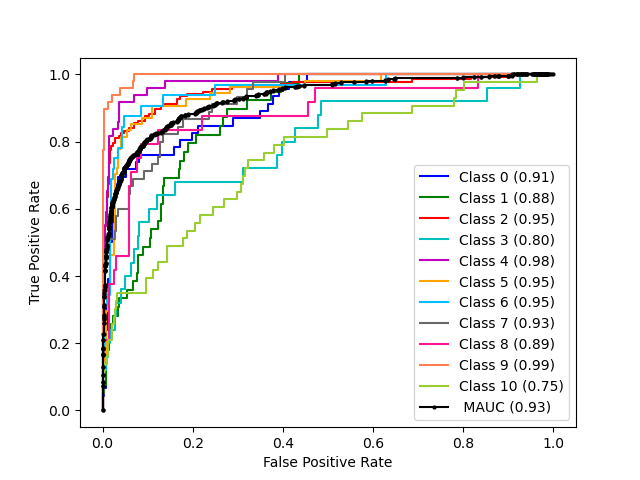}
        \includegraphics[width=0.32\linewidth]{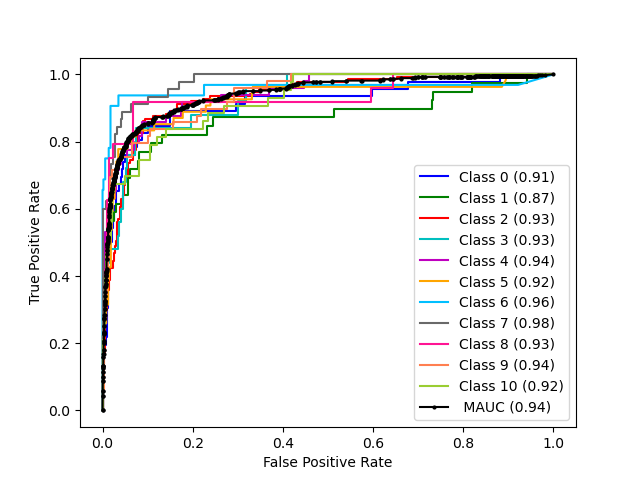}
        \includegraphics[width=0.32\linewidth]{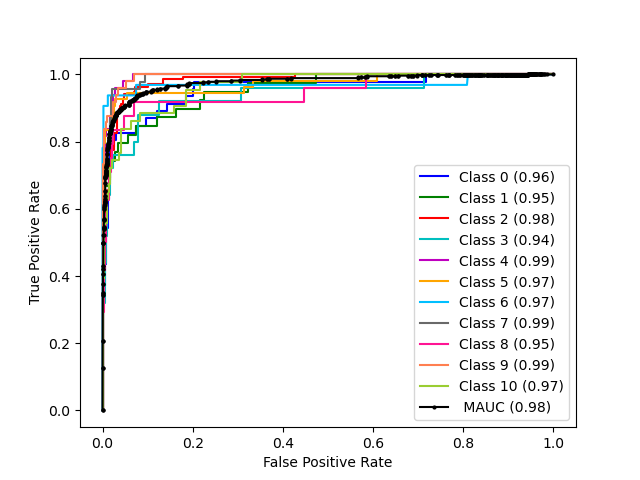}
    \caption{We report the ROC curves and the micro-average scores for the best performing models using different types of genetic information: (a) Using the presence of mutation in a gene as a feature for top 15,000 genes, (b) using CNV information in a gene as a feature for top 17,000 genes, (c) using both encoded SNV and CNV features with top 34,000 features.}
    \label{fig:MAUC_all}
\end{figure*}
\begin{table*}[]
\centering
\caption{Performance of cancer prediction model for each for 11 classes. For each machine learning model and feature selection combination, we report the model with highest performance. The best performing model among them, according to test accuracy, is in boldface. Feature selection denotes the statistical feature selection. Accuracy denoted here is the test accuracy.}
\label{tab:all_results}
\begin{tabular}{cccccccccc}
 & \multicolumn{1}{c}{Feature type} &  \#features & Model type & Feature selection & Accuracy & MAUC & Precision & Recall & F-score \\
 \hline
 \hline
 & SNV presence & 15,000 & Logistic regression & $\chi^2$ & 66.85 & 0.928 & 0.670 & 0.668 & 0.660 \\
 & CNV only & 17,000 & Logistic regression & $\chi^2$ & 71.27 & 0.940 & 0.718 & 0.712 & 0.711 \\
 & CNV + encoded SNV & 13,000 & SVM (linear) & $\chi^2$ & 68.13 & 0.942 & 0.685 & 0.681 & 0.680 \\
 & CNV + encoded SNV & 37,000 & SVM (rbf) & $\chi^2$ & 64.82 & 0.949 & 0.685 & 0.648 & 0.634 \\
 & CNV + encoded SNV & 34,000 & SVM (polynomial) & $\chi^2$ & 69.98 & 0.954 & 0.712 & 0.699 & 0.698 \\
 & CNV + encoded SNV & 43,000 & DNN & $\chi^2$ & 63.16 & 0.925 & 0.658 & 0.631 & 0.628 \\
 & \textbf{CNV + encoded SNV} & \textbf{34,000} & \textbf{Logistic regression} & \textbf{$\chi^2$} & \textbf{83.61} & \textbf{0.976} & \textbf{0.834} & \textbf{0.836} & \textbf{0.834} \\
 & CNV + encoded SNV & 32,000 & Logistic regression & MI & 81.95 & 0.972 & 0.822 & 0.819 & 0.818 \\
 & CNV + encoded SNV & 40,000 & Logistic regression & f-score & 82.68 & 0.974 & 0.827 & 0.826 & 0.824\\
  \hline
\end{tabular}
\end{table*}

\begin{table}[t]
\centering
\caption{Percentage of samples belonging to a class and the corresponding test accuracy of the class }
\label{tab:classwise}
\begin{tabular}{lcc}
 & \begin{tabular}[c]{@{}c@{}}Percent of \\ samples\end{tabular} & Accuracy \\
 \hline
\hline
Class 0 (Bladder) & 9.76 & 78.26 \\
Class 1 (Breast) & 7.46 & 74.35 \\
Class 2 (Bronchus and lung) & 23.08 & 91.24 \\
Class 3 (Cervix uteri) & 5.71 & 64.0 \\
Class 4 (Colon) & 9.53 & 87.75 \\
Class 5 (Corpus uteri) & 7.6 & 85.18 \\
Class 6 (Kidney) & 5.39 & 90.625 \\
\begin{tabular}[c]{@{}l@{}}Class 7 (Liver and intrahepatic \\ bile ducts)\end{tabular} & 6.63 & 93.33 \\
Class 8 (Ovary) & 5.85 & 70.83 \\
Class 9 (Skin) & 9.44 & 87.75 \\
Class 10 (Stomach) & 9.49 & 65.11\\
\hline
\end{tabular}
\end{table}

\section{Experimental evaluation}\label{s:results}
The evaluation of private cancer prediction methodology is dependent on the precision of the plaintext model and performance of the private inference protocol. We report the evaluation of somatic mutation encoding towards accurate cancer prediction in plaintext in subsection \ref{ss:results_part1} and the performance of the ML model (based on our matrix multiplication methodology) in subsection \ref{ss:results_part2}.

\subsection{Evaluation of plaintext cancer prediction}\label{ss:results_part1}
\subsubsection{Using only presence of an SNV on a gene}
We report different performance metrics and the information on the features used for different models tested in Table \ref{tab:all_results}.
We observe that our model achieves a test accuracy of $66.85\%$ and a micro-average area under curve of $0.928$ with top 15,000 features. We also plot an Receiver Operating Characteristics (ROC) curve (Fig.~\ref{fig:MAUC_all}) for each class and observe that skin cancer (class 9) detection has the highest area under the curve of $0.994$ while stomach cancer (class 10) prediction has the lowest area under the curve of $0.754$. Although on a slightly different dataset, other ML-based cancer prediction achieved a similar test accuracy of $65.5\%$ \cite{deep_gene} and of $70.08\%$ \cite{gdl}. These methods also used the SNV frequency to prune the number of features.

\subsubsection{Using only CNV of the genes}\label{ss:cnv_results}
We also experiment with just the copy number information for all the 25,128 genes and run a $\chi^2$ test to select the top genes. 
We achieve a slightly higher test accuracy of $71.27\%$ with 17,000 features. From Fig.~\ref{fig:MAUC_all}, we observe that the micro average area also improves to $0.94$ with the lowest area under curve for detection of breast cancer (class 1) at $0.87$ (from $0.754$). 
Higher test accuracy and MAUC show that CNVs have more distinguishing power on the type of cancers than SNVs, when considered individually. 
\begin{table*}[t]
\centering
\caption{Performance of individual models for each cancer type in terms of test accuracy and micro-average area under curve.}
\begin{tabular}{lccccc}
 & Accuracy & MAUC & Precision & Recall & F-score \\
 \hline
 \hline
Class 0 (Bladder) & 95.58 & 0.982 & 0.952 & 0.955 & 0.952 \\
Class 1 (Breast) & 94.65 & 0.983 & 0.942 & 0.946 & 0.944 \\
Class 2 (Bronchus and lung) & 91.34 & 0.974 & 0.911 & 0.913 & 0.912 \\
Class 3 (Cervix uteri) & 97.42 & 0.989 & 0.972 & 0.974 & 0.973 \\
Class 4 (Corpus uteri) & 97.42 & 0.997 & 0.974 & 0.974 & 0.974 \\
Class 5 (Colon) & 96.86 & 0.991 & 0.967 & 0.968 & 0.968 \\
Class 6 (Kidney) & 97.97 & 0.996 & 0.979 & 0.979 & 0.979 \\
\begin{tabular}[c]{@{}l@{}}Class 7 (Liver and \\ intrahepatic bile duct)\end{tabular} & 95.39 & 0.992 & 0.952 & 0.953 & 0.953 \\
Class 8 (Ovary) & 97.60 & 0.993 & 0.974 & 0.976 & 0.975 \\
Class 9 (Skin) & 97.42 & 0.995 & 0.973 & 0.974 & 0.973 \\
Class 10 (Stomach) & 96.13 & 0.982 & 0.958 & 0.961 & 0.958 \\
\hline
\end{tabular}
\label{tab:binary}
\end{table*}
\subsubsection{Final model: Using both SNV and CNV information}
We select top $n$ features using both SNV and CNV information as described above and train several models to evaluate different machine learning algorithms for the tumor classification task. SVM with linear, rbf, and polynomial kernels achieve the best test accuracy of $68.13\%$, $64.82\%$, and $69.98\%$ with 13,000, 37,000, and 34,000 features respectively. Therefore, SVM does not show much improvement from our baseline that used only the presence of SNVs as features. 
Logistic regression model with Lasso penalty shows the best performance across all performance metrics. We achieve a test accuracy of $83.61\%$ with 34,000 features thereby also reducing the number of features. We also observe an improvement in the micro average area to $0.976$ when compared with models using just CNV or just presence of SNV. ROC for individual classes also have higher scores with the lowest area under curve of $0.94$ for cervical cancer (class 3). All the classes achieve an ROC area under curve of more than $0.9$. This experiment also shows that although CNVs are more informative, using both CNV and SNVs result in the highest prediction accuracy. We also test our logistic regression model using mutual information and f-score as feature selection methods but we achieve lower test accuracy of $81.95\%$ with 32,000 features, and $82.68\%$ with 40,000 features, respectively (Table\ref{tab:all_results}). 
 
\subsubsection{Binary models}
We built 11 specific models for each cancer type. We use both CNV and SNV information to select the top 34,000 features. In these experiments, we train 11 different models, where each one detects a particular tumor. We report the performance (test accuracy and micro-average score) in Table \ref{tab:binary}. We can see that all of the individual models have a test accuracy of more than $90\%$ and a micro-average area under curve of more than $0.9$. The best performing classifier is for kidney with a test accuracy of $97.97\%$ and a micro-average score of $0.996$. Even the worst performing binary model (for bronchus and lung) has a test accuracy of $91.34\%$ and micro-average score of $0.974$. Binary models have also been evaluated in other somatic genetic information-based cancer classification tasks \cite{gdl} where the authors also achieve high performance for individual tasks but suffer in performance in cancer prediction using all labels.

\subsubsection{Predictive genes}\label{ss:predictive}
\begin{figure}[t]
    \centering
    \includegraphics[scale=0.48]{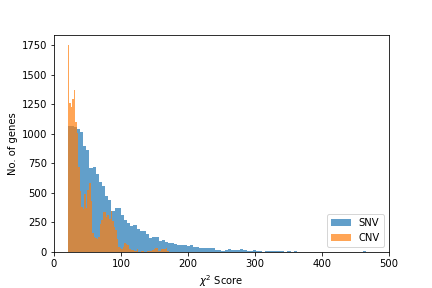}
    \caption{The distribution of feature scores for genes contributing CNV and SNV data in the final model.}
    \label{fig:feature_imp}
\end{figure}

We discuss our findings on the top genes selected. 17,962 genes are selected for CNV data and 16,038 are selected for SNV data as the most informative features. However, more than $60\%$ of these genes are common, i.e. 11,133 genes are selected based on both CNV and SNV information. Considering the $\chi^2$ scores, we also observe that the top 1,030 genes come from using SNV information. We plot histograms of $\chi^2$ statistic scores in Fig.~\ref{fig:feature_imp} and observe that the genes selected based on SNVs is more flat i.e. there are more genes with higher scores. We also verified that the highest score of a gene selected based on SNVs is $\approx 10 \times$ higher than the highest value of a gene selected based on CNVs. Therefore, from feature selection perspective using $\chi^2$ test, SNV data on the selected genes are statistically \textit{more important} than their CNV counterparts. But CNV information improves the test accuracy by adding potentially more relevant biological information. 

When we investigated the top 10 most informative genes based on the SNV information (Table \ref{tab:top_genes}), we found ``PTEN" are ``APC" genes, which are known tumor suppressors; ``MUC16" gene, which is a biomarker for ovarian cancer; ``ZFHX3" gene, which is implicated in prostate cancer; ``CCDC168" gene, which is known to be associated with Prostate Carcinoma and Uterine Body Mixed Cancer. The other 5 genes in this list are also implicated in important cellular activities that could potentially be related to cancer. These genes and their corresponding Gene Ontology (GO) term enrichment are depicted in Table \ref{tab:top_genes}.

\begin{table}[]
\caption{Genes selected by our model. The left column denotes the selected genes and the right column represents their GO enrichment terms.}
\label{tab:top_genes}
\begin{tabular}{ll}
\begin{tabular}[c]{@{}l@{}}CNV-based \\ top 10 genes\end{tabular} & Go enrichment analysis \\
\hline
\hline
RB1 & \begin{tabular}[c]{@{}l@{}}DNA-binding transcription factor activity and \\ enzyme binding\end{tabular} \\
CDKN2A & transcription factor binding \\
LINC00441 & NA \\
DGKH & \begin{tabular}[c]{@{}l@{}}NAD+ kinase activity and diacylglycerol kinase \\ activity\end{tabular} \\
RCBTB2 & Ran guanyl-nucleotide exchange factor activity \\
CDKN2B-AS1 & NA \&  Intracranial Aneurysm and Periodontitis \\
LPAR6 & G protein-coupled receptor activity \\
AKAP11 & \begin{tabular}[c]{@{}l@{}}protein kinase A binding and protein phosphatase \\ 1 binding\end{tabular} \\
CDKN2B & \begin{tabular}[c]{@{}l@{}}protein kinase binding and cyclin-dependent \\ protein serine/threonine kinase inhibitor activity\end{tabular} \\
ITM2B & amyloid-beta binding \\
\hline
\begin{tabular}[c]{@{}l@{}}SNV-based\\ top 10 genes\end{tabular} & Go enrichment analysis \\
\hline
\hline
TTN & nucleic acid binding and identical protein binding \\
PTEN & protein kinase binding and magnesium ion binding \\
APC & microtubule binding \\
MUC16 & metabolism \\
DST & calcium ion binding and actin binding \\
ZFHX3 & \begin{tabular}[c]{@{}l@{}}nucleic acid binding and sequence-specific DNA \\ binding\end{tabular} \\
CCDC168 & NA \\
ATRX & chromatin binding and helicase activity \\
DNAH5 & ATPase activity and microtubule motor activity \\
PIK3R1 & GTP binding and transcription factor binding \\
\hline
\end{tabular}
\end{table}
The first gene in the top 10 most informative genes based on the CNV information (Table \ref{tab:top_genes}) is the first ever known tumor suppressor ``RB1". Similarly, ``CDKN2A" and ``CDKN2B" genes are also known tumor suppressors. ``RCBTB2" gene is known to be repressed in prostate cancer. The ``CDKN2B-AS1" gene has the silencing power of many other genes in the genome and strongly implicated in various cancer types. The other 5 genes in this list are also implicated in important cellular activities that could potentially be related to cancer. 

\subsubsection{Insights}The first insight is that biological intuition combined with high-dimensional data analysis methods can together achieve high accuracy (MAUC) while reducing the effective number of features. We further validate, using other studies, that our ML model indeed selects features that are biologically relevant. Using domain expertise, our ML model achieved a high performance of 0.98 MAUC with logistic regression ML model.

\begin{figure}
    \centering
    \includegraphics[scale=0.5]{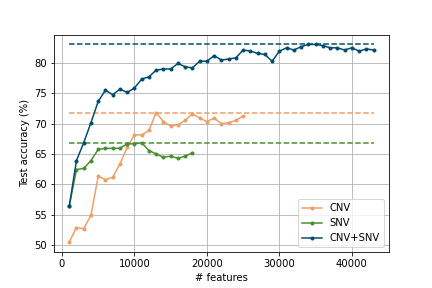}
    \caption{Variation of the achieved test accuracy of the trained models with different number of features selected using $\chi^2$ test.}
    \label{fig:acc_vs_features}
\end{figure}

We further plot the variation of test accuracy (of a model trained using best hyper-parameters) with the number of features in Fig. \ref{fig:acc_vs_features} and observe that the test accuracy clearly rises when the number of features is lower than the number of samples and begins to saturate only after 30K features. 
From Fig. \ref{fig:acc_vs_features} we see that to achieve a test accuracy of more than $80\%$, we need at least 20K features, and to achieve the highest test accuracy, we need more than 30K features. Therefore, private genomic analysis is one such application where extremely high-dimensional must be processed. The second insight motivates the development of matrix multiplication algorithm, which when implemented using BFV, can result in fast yet private cancer prediction.

\subsection{Privacy-preserving model evaluation}\label{ss:results_part2}
We evaluate our privacy-preserving cancer prediction model on AMD Ryzen Threadripper 3960X 24-Core Processor with 128 GB RAM using 24 threads running Ubuntu 20.04 LTS. Encryption and computation operations are threaded, while decryption runs on single core. We implement our model using the E3 framework \cite{e3} with the underlying Microsoft SEAL library \cite{seal} and encryption parameters set as: polynomial degree $n = 8192$, and plaintext moduli $t_0 = 1073872897$ and $t_1 = 114689$, with a required security level of 128-bits.
The cancer prediction model is hosted in the server and the client sends the encrypted genomic data to the server. As a use case, we privately compute cancer label for 543 patients, which constitutes $20\%$ of the dataset. We compare our private logistic regression model with private logistic regression model implemented using standard matrix multiplication (dubbed as standard LR). Please note all private models are implemented using BFV scheme with E3 framework.

\subsubsection{Timing evaluation}
\begin{figure}
    \centering
    \includegraphics[scale=0.5]{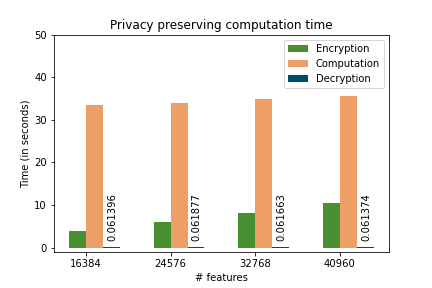}
    \caption{Performance of our private LR-based cancer prediction model as a function of features.}
    \label{fig:results}

\end{figure}

We report the encryption, decryption, and computation time required for private cancer prediction in Fig. \ref{fig:results}. 
The time taken to calculate the final cancer label, which is effectively the result of matrix multiplication $(Wx_{test}+b)$, is denoted by computation. Computation, understandably, is the most costly operation in private cancer prediction. We observe that even if the number of features increase from 16K to more than 40K ($2.5\times$), the computation time only increases from 33.44 seconds to 35.52 seconds ($1.06\times$), which corresponds to $\approx7\%$ increase in test accuracy. Therefore, the matrix multiplication is not the bottleneck for private cancer prediction. The time needed for encryption of the test samples increases with the number of features, with 3.87 seconds for 16K features to 10.40 seconds for 40K features ($2.68\times $) which indicates a linear increase in the encryption time as a function of number of features. Decryption is the least expensive operation (less than 1 second) as compared to encryption and computation; the values for decryption time are labelled in Fig. \ref{fig:results}. The maximum total time for private inference of the entire test dataset is required when processing 40,960 features, and is 46.77 seconds.

\subsubsection{Latency}

\begin{figure}
    \centering
    \includegraphics[scale=0.5]{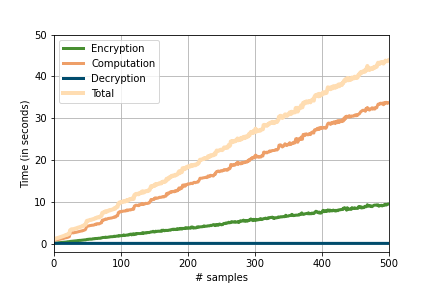}
    \caption{Timing for different operations as a function of number of test samples.}
    \label{fig:latency}

\end{figure}

As mentioned in section \ref{s:introduction} private computations using HE are generally designed for high throughput, since popular FHE schemes support batching. For our application, we also prioritize latency, i.e. evaluation of a single sample. We report our findings in Fig. \ref{fig:latency}. From the figure we observe that the total amount of time to privately compute cancer label for a sample is 1.08 seconds and there is a linear increase in time with the number of samples.

\begin{figure}[t]
    \centering
    \includegraphics[scale=0.5]{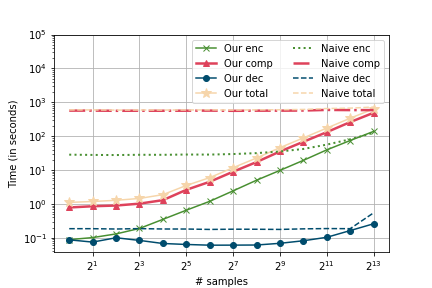}
    \caption{Latency comparison of our matrix multiplication algorithm with standard privacy-preserving matrix multiplication.}
    \label{fig:latency_comparison}
\end{figure}

\subsubsection{Comparison to standard LR}
In order to accurately quantify the performance benefits of our proposed methodology, we implement a privacy-preserving version of standard matrix multiplication using BFV. For the experiment we generate synthetic data in the same format (CNVs and SNVs) for 8192 individuals and measure time for encryption, computation, decryption operations. We plot the timing results in a log-log graph in Fig. \ref{fig:latency_comparison}. We observe that the total time required for private inference implemented using standard matrix multiplication for similar number of individuals as the test set is approximately 10 minutes, approximately $10\times$ more than our methodology. Also, the total time required for private inference on 1 individual is 598.25 seconds (similar time required for thousands of individuals), which is $550\times$ more than the time required by our algorithm. Therefore, as compared to standard matrix multiplication, commonly used for implementation of ML models, our algorithm has lower latency, and higher throughput.

\subsubsection{Generalizing high-dimensional private inference} Healthcare models are difficult to port trivially across datasets (as discussed in section \ref{s:introduction}). Cancer detection ML model is no exception. However, our matrix multiplication algorithm is not dependent on input data or weight values (like quantization-based DNN design techniques \cite{qnn}) and thus, can be reused for datasets requiring HE-based high-dimensional inference. The transferability of our private inference algorithm across applications is an added advantage.

\section{Conclusion}\label{s:conclusion}
Current solutions for HE-based privacy preserving inference suffer from impractical overheads; which are further aggravated when dealing with high-dimensional genomic data.  
In this work we develop a solution for privacy preserving cancer inference on genomic data. We first leverage biological intuition to structure the mutation data and reduce the dimensionality to a practicable limit. For our privacy preserving ML model, we propose a matrix multiplication algorithm to implement logistic regression model, optimized for high throughput and low latency. Our analysis on a real-world genomic dataset shows that our solution achieves cancer prediction MAUC of 0.98 on test dataset and can be computed on encrypted genomic data at $\approx$ 1 second/patient. 
\printbibliography

\section*{Appendix}\label{s:appendix}
Here we describe the matrix multiplication of $Y=\hat{X} \times \bar{W}$, where $\hat{X}$ is the encrypted input matrix (encoded genomic data) and  $\bar{W}$ is the encoded matrix of LR weights. The polynomial degree is $n$, $|X|$ is the number of inputs, $|Y|$ is the number of outputs, and $f$ is the number of features. The operator $\times$ stands for the standard matrix multiplication, while $\otimes$ represents our algorithm, $[\cdot]_n$ is modular reduction over $n$, and the intervals $[a,b)$ and $[a,b]$ represent elements packed in a ciphertext. When $b<a$, there is a rotation of the $n$ elements of the ciphertext. Function $\rho(\cdot)$ is the element-wise addition of all rotations of a ciphertext, and function $\alpha(\cdot)$ represents the compression part of the algorithm, where one slot of $n$ ciphertexts is selected and combined into a new ciphertext.
\input{matrix}
\end{document}

%% file: intro2.tex
\section{Introduction}\label{s:introduction}
The improvement in computing power and an overwhelming wealth of data have enabled the viability of statistical learning methods and have achieved unprecedented performance in  several fields of science, engineering, and finance. Artificial Intelligence (AI)-based techniques are also being incorporated by federal agencies to make critical infrastructure safe, secure, and robust~\cite{ai_secure} and healthcare critical infrastructure is not an exception. From first generation rule-based healthcare AIs, the sector is slowly moving towards ML-based methods to investigate, understand, and use complex data patterns for clinical diagnosis~\cite{ai_health}. ML-based healthcare predictions are currently far from having generalizable diagnostic abilities. For example, trained ML models for disease prediction are extremely data dependent, and cannot be seamlessly transferred as in the case of image classification tasks. One such disease prediction application, with profound impact, is cancer (tumor) detection. An early/fast diagnosis of cancer (cancer inference) by a trained ML model can save precious time in prognosis and treatment of a patient.

Precision medicine is the process of tailoring the right diagnosis and treatment for the right person at the right place. One of the biggest components of precision medicine is to incorporate patients' genetic information to the diagnosis, treatment, and decision making \cite{precision}. In the context of precision cancer medicine, distinguishability between genetic mutations of normal and malignant tissues is the crux of cancer genomics. These genetic changes developed during a person's life in malignant tumor cells are called somatic (or acquired) changes and are accountable for more than $90\%$ of cancer cases. Somatic Single-Nucleotide Variation (SNV) and Copy-Number Variation (CNV) on protein-coding genes, especially on oncogenes, tumor suppressors and cell cycle regulators are known to cause tumor formation and progress. However, the heterogeneity in various levels makes it difficult to understand precisely which gene is involved in which cancer type. Somatic mutation rate is different across cancer types; even within a single type, this rate is different across patients~\cite{mut_rate}. Conversely, origins of distinct cancer types have been found to share similarities, thus making distinguishability even harder~\cite{similar_origin}. It is important to find explainable relationships between the somatic mutations, the genes that they affect and the type of cancer. We explore the problem of cancer genomics using a real-world dataset consisting of more than 2 million CNV and SNV information of 11 different cancer types~\cite{idash20}.

While the use of genomics in cancer detection seems promising for comprehensive understanding of the disease and its treatment, there are major privacy-related concerns. Genomic data is extremely characterizing and identifying, and the data may pin-point to the exact patient. It is also permanent and cannot be changed like other private data (passwords, credit cards, etc.). A partial leak of genomic data may reveal important information about the individual and may also be used to reconstruct their genome \cite{genome_attack}. A patient, waiting for their cancer diagnosis using a server-hosted state-of-the-art ML-based predictive models, should not be subjected to such privacy risks. 

High privacy risks of genomic data leakage calls for data to be always encrypted during inference. Homomorphic Encryption (HE) is an encryption scheme which allows for encrypted computation, i.e. the encrypted genomic data can be sent to the ML model in the server and the patient can receive the encrypted diagnosis.
The first Fully HE (FHE) scheme for arbitrary computation proposed by Gentry et. al. \cite{gentry} had prohibitive computational overheads for real-world applications like ML-based inference. Since the non-linear activation functions were the major bottleneck for private ML-based inference, researchers mainly focused on approximating the non-linear function using square function \cite{cryptonets}, first terms of Taylor expansion \cite{access_imputation}, or piece-wise linearization \cite{minionn}, for fast private inference. A possible solution for this problem is to bypass the non-linear function using small models like logistic regression (as we show in section \ref{sss:method_approx}). But inherent high-dimensionality of genomic data makes the encrypted linear operation impractical for privacy preserving genomics. Using an ML model on the raw cancer dataset would require multiplication of matrices with millions of columns, which is extremely expensive in HE.
The intuitive solution towards faster private inference is to reduce the number of computations. This translates to reducing the number of features using feature selection, as can be seen in other HE applications in genomics \cite{access_imputation,ultrafast}.

For dimensionality reduction, we develop a feature engineering methodology involving feature (gene) selection and genetic (mutation) information encoding, and is based on a combination of biological intuition and statistical tests. Trivially using statistical scores may result in overfitting especially for genomic datasets with the number of predictors $(f)$ several times larger than the number of samples $(|X|)$ i.e. $f>>|X|$.
Domain knowledge is increasingly being used for feature engineering in healthcare predictive models \cite{domain_healthcare} because such methods are not only more interpretable but also integrate years of medical research and case-studies. Healthcare ML models need to be interpretable for sustainability \cite{healthcare_explainability}. Therefore, it is sustainable to use explainable models (like SVM, logistic regression, etc.) than Deep Neural Networks (DNN) that are essentially a black-box. This practically eliminates the automatic feature engineering capabilities of DNNs. Using our methodology of somatic mutation encoding, we reduce the dimensionality of the task (from over 2 million mutations to 43K features), but still the genomic data remains high-dimensional as compared to benchmark ML datasets (like MNIST with image size of $28\times28=784$ features or CIFAR with image size of $32\times32\times3=3072$ features). Since our application \textit{needs} several thousands of features for accurate predictions, not only our time budget is completely exhausted by the linear operation, but also standard matrix multiplication does not offer the performance needed.
Another drawback of current HE-based implementations of private inference is that they are designed to maximize throughput, computing on thousands of inputs together to improve efficiency in a cumulative way. However, they suffer in latency, i.e. the algorithms would take the same time to compute on just one input as it would take for thousands of inputs.
But the real-world application we consider in this work benefits from improved latency, as it a common use case to have to analyze the genome of only a single patient and not wait for thousands of patients to have their tests done. In summary, to enable practical real-world private inference, we need the ability to compute on high-dimensional data in the encrypted domain with low latency and high throughput.


%% file: algorithm.tex
\begin{algorithm}[t!]
    \small
    \caption{Proposed matrix multiplication algorithm}\label{alg:mm}
    \begin{algorithmic}[1]
        \Function{matrixMultiplication}{$\hat{X}$, $\bar{W}$, $n$}
            \State vector$<$Ciphertext$>$ $\hat{R_0}$($|\hat{X}|$)
            \State Plaintext $\bar{p}$ = encode( vector$<$integer$>$($n, 0$) )
            \State $l$ = $0$
            \For{( $i=0$; $i<|\hat{X}|$; $i$++ )}
                \For{( $j=0$; $j<|\bar{W}|$; $j$++ )}
                    \State // multiplication
                    \State vector$<$Ciphertext$>$ $\hat{\tau}$
                    \For{( $k=0$; $k<\ceil{n/f}$; $k$++ )}
                        \State $\hat{\tau}$.append($\hat{X}[i][k] \cdot \bar{W}[j][k]$)
                    \EndFor
                    \State // addition
                    \State Ciphertext $\hat{c}$ = add\_many($\hat{\tau}$)
                    \For{( $s=n>>1$; $s > 0$; )}
                        \State $s >>= 1$
                        \State $\hat{c}$ += $\hat{c} << s$ // rotate and accumulate
                    \EndFor
                    \State // masking
                    \State $\bar{p}$.encodeAt($l, 0$)
                    \State $l = i \cdot |\bar{W}| + j \mod n $
                    \State p.encodeAt($l, 1$)
                    \State $\hat{R_0}$[$i$].append($\hat{c} \cdot \bar{p}$)
                \EndFor
            \EndFor
            \State // compression
            \State vector$<$Ciphertext$>$ $\hat{R_1}, \hat{Y}$
            \State $l$ = $0$
            \For{( $\hat{V}$ \textbf{in} ${\hat{R_0}}$ )}
                \For{( $\hat{c}$ \textbf{in} ${\hat{V}}$ )}
                    \If{( $l$ == $0$ )}
                        \If{( $|\hat{R_1}|$ != $0$ )}
                            \State $\hat{Y}$.append( add\_many($\hat{R_1}$) )
                            \State $\hat{R_1}$.clear()
                        \EndIf
                    \EndIf
                    \State $\hat{R_1}$.append($\hat{c}$)
                    \State $l$ = $l + 1 \mod n$
                \EndFor
            \EndFor
            \If{( $|\hat{R_1}|$ != $0$ )}
                \State $\hat{Y}$.append( add\_many($\hat{R_1}$) )
            \EndIf
            \Return $\hat{Y}$
        \EndFunction
    \end{algorithmic}
\end{algorithm}
\setlength{\textfloatsep}{0pt}

%% file: matrix.tex
\begin{align*}
    Y =
    \begin{bmatrix}
        x_{0,0}     & \hdots & x_{0,f-1}     \\
        \vdots      & \ddots & \vdots        \\
        x_{|X|-1,0} & \hdots & x_{|X|-1,f-1}
    \end{bmatrix}
    \times
    \begin{bmatrix}
        w_{0,0}     & \hdots & w_{0,|Y|-1} \\
        \vdots      & \ddots & \vdots      \\
        w_{f-1,0} & \hdots & w_{f-1,|Y|-1}
    \end{bmatrix}
    = \\
    \begin{bmatrix}
        x_{0,0}     & \hdots & x_{0,f-1}     \\
        \vdots      & \ddots & \vdots        \\
        x_{|X|-1,0} & \hdots & x_{|X|-1,f-1}
    \end{bmatrix}
    \times
    \begin{bmatrix}
        w_{0,0}     & \hdots & w_{0,f-1}   \\
        \vdots      & \ddots & \vdots      \\
        w_{|Y|-1,0} & \hdots & w_{|Y|-1,f-1}
    \end{bmatrix}^{-1}
    = \\
    \begin{bmatrix}
        x_{0,[0,n)}     & \hdots & x_{0,[f-n,f)}    \\
        \vdots          & \ddots & \vdots           \\
        x_{|X|-1,[0,n)} & \hdots & x_{|X|-1,[f-n,f)}
    \end{bmatrix}
    \otimes
    \begin{bmatrix}
        w_{0,[0,n)}     & \hdots & w_{0,[f-n,f)}    \\
        \vdots          & \ddots & \vdots           \\
        w_{|Y|-1,[0,n)} & \hdots & w_{|Y|-1,[f-n,f)}
    \end{bmatrix}
    = \\
    \alpha\Bigg(\rho\Bigg(\begin{bmatrix}
        \sum_{i=0}^{\ceil{f/n}-1}{x_{0,[i \cdot n,(i+1) \cdot n)} \cdot w_{0,[i \cdot n,(i+1) \cdot n)}}     & \hdots & \sum_{i=0}^{\ceil{f/n}-1}{x_{0,[i \cdot n,(i+1) \cdot n)} \cdot w_{|Y|-1,[i \cdot n,(i+1) \cdot n)}} \\
        \vdots                                                                                               & \ddots & \vdots                                                                  \\
        \sum_{i=0}^{\ceil{f/n}-1}{x_{|X|-1,[i \cdot n,(i+1) \cdot n)} \cdot w_{0,[i \cdot n,(i+1) \cdot n)}} & \hdots & \sum_{i=0}^{\ceil{f/n}-1}{x_{|X|-1,[i \cdot n,(i+1) \cdot n)} \cdot w_{|Y|-1,[i \cdot n,(i+1) \cdot n)}}
    \end{bmatrix}\Bigg)\Bigg)
    = \\
    \alpha\Bigg(\rho\Bigg(\begin{bmatrix}
        t_{0,[0,n)}     & \hdots & t_{0,[|Y|-n,|Y|)}    \\
        \vdots          & \ddots & \vdots               \\
        t_{|X|-1,[0,n)} & \hdots & t_{|X|-1,[|Y|-n,|Y|)}
    \end{bmatrix}\Bigg)\Bigg)
    = \\
    \alpha\Bigg(\begin{bmatrix}
        \sum_{i=0}^{n-1}{t_{0,[i,[n+i]_n]}}     & \hdots & \sum_{i=0}^{n-1}t_{0,[|Y|-n+i,|Y|-n+[|Y|-1+i]_n]} \\
        \vdots                                  & \ddots & \vdots                                            \\
        \sum_{i=0}^{n-1}{t_{|X|-1,[i,[n+i]_n]}} & \hdots & \sum_{i=0}^{n-1}{t_{|X|-1,[|Y|-n+i,|Y|-n+[|Y|-1+i]_n]}}
    \end{bmatrix}\Bigg)
    = \\
    \alpha\Bigg(\begin{bmatrix}
        u_{0,[0,n)}     & \hdots & u_{0,[|Y|-n,|Y|)} \\
        \vdots          & \ddots & \vdots            \\
        u_{|X|-1,[0,n)} & \hdots & u_{|X|-1,[|Y|-n,|Y|)}
    \end{bmatrix}\Bigg)
    = \\
    \begin{bmatrix}
        \{ u_{0,0}, u_{0,n}, u_{0,2n}, \hdots, u_{0,(n-1) \cdot n} \}    & \hdots & \{ u_{0,|Y|-(n-1) \cdot n -1}, \hdots, u_{0,2n-1}, u_{0,|Y|-n-1}, u_{0,|Y|-1} \} \\
        \vdots          & \ddots & \vdots            \\
        \{ u_{|X|-1,0}, u_{|X|-1,n}, \hdots, u_{|X|-1,(n-1) \cdot n} \}    & \hdots & \{ u_{|X|-1,|Y|-(n-1) \cdot n -1}, \hdots, u_{|X|-1,|Y|-n-1}, u_{|X|-1,|Y|-1} \}
    \end{bmatrix}
    = \\
    \begin{bmatrix}
        y_{0,[0,n)}     & \hdots & y_{0,[|Y|-n,|Y|)} \\
        \vdots          & \ddots & \vdots            \\
        y_{|X|-1,[0,n)} & \hdots & y_{|X|-1,[|Y|-n,|Y|)}
    \end{bmatrix}
\end{align*}
\normalsize